\documentclass[12pt]{preprint}
\usepackage[dvips]{graphicx}
\usepackage{amsmath}
\usepackage{amsfonts}
\usepackage{amssymb}
\usepackage{times}
\usepackage{sub_JP}
\usepackage{MHequ}

\def\OO{{\mathcal O}}

\def\L{{\rm L}}
\def\R{{\rm R}}

\def\kB{k_{\rm B}}

\def\eref#1{(\ref{#1})}
\let\rho=\varrho

\def\sref#1{Sect.~\ref{#1}}
\def\fref#1{Fig.~\ref{#1}}

\begin{document}

\title{Superdiffusive Heat Transport in a class of Deterministic One-Dimensional Many-Particle Lorentz gases}
\author{Pierre Collet${}^1$, Jean-Pierre Eckmann${}^{2,3}$ and Carlos~Mej\'{\i}a-Monasterio${}^4$
}
\institute{
${}^1$Centre de Physique Th\'eorique, CNRS UMR 7644, Ecole Polytechnique,
 F-91128 Palaiseau Cedex (France)\\
${}^2$D\'epartement de Physique Th\'eorique, Universit\'e de Gen\`eve\\
${}^3$Section de Math\'ematiques, Universit\'e de Gen\`eve\\
${}^4$Istituto dei Sistemi Complessi, Consiglio Nazionale delle Ricerche, Sesto Fiorentino Italy
}
\maketitle

\begin{abstract}
We study heat transport in  a one-dimensional chain of a finite number $N$
of  identical  cells,  coupled  at  its boundaries  to  stochastic  particle
reservoirs.  At  the center  of each cell,  tracer particles  collide with
fixed scatterers,  exchanging momentum.
In a  recent paper,
\cite{CE08}, a spatially continuous  version of  this model was
derived in a scaling regime where the scattering probability of the tracers is  $\gamma\sim1/N$, corresponding to the 
Grad limit.
A  Boltzmann  type  equation
describing the transport of heat  was obtained.  In this paper, we  show 
numerically that  the Boltzmann
description  obtained in
\cite{CE08} is indeed a bona fide  limit of the particle model. 
Furthermore, we also study the heat
transport of the  model when the scattering probability is one,
corresponding to deterministic dynamics.
At a coarse grained level the  model behaves as a 
persistent random walker  with a broad  waiting time distribution and  
strong correlations
associated to  the deterministic  scattering.  We show, that, in  spite 
of the  absence of
global conserved quantities, the model leads to a superdiffusive heat 
transport.
\end{abstract}
\thispagestyle{empty}
\section{Introduction}
\label{sec:intro}
The rigorous  description of (classical) non-equilibrium  steady states (NESS)
remains an elusive  problem, despite some remarkable progress  in the last few
years,  as described,  {\it e.g.},  in  \cite{BLRB00,LLP03}. One  of the  main
reasons seems to be the impossibility  to guess the stationary state, which is
one of the  mechanisms which work in equilibrium  statistical mechanics. There
are therefore many studies which try to  understand better what the
essentials of  NESS are, how different  models fit together, and  what are the
best  descriptions of  NESS. Among  the  few works  where the  NESS have  been
obtained,  we  mention the  harmonic  chain  coupled  to Langevin  heat  baths
\cite{RLL67}  and   anharmonic  chains  coupled  to   infinite  (non  compact)
reservoirs \cite{EPRB99,bc2007}.

In this paper, we study heat  and particle conduction in a 1-dimensional model
introduced  in \cite{CE08},  and which  is  a variant  of a  model studied  in
\cite{MMLL01, EY06}.  We analyze several  of its properties and  in particular
show, by numerical study, that the Boltzmann description of the model obtained
in \cite{CE08}  is indeed  a bona  fide limit of  the particle  models studied
here, but only so  if the coupling, per site, is of  order $\OO(1/N)$ when the
number of sites  is (large) $N$. The  model does {\emph not} seem  to obey the
Fourier law. The reader familiar with transport problems will also notice that
(due  to the  strictly 1-dimensional  character  of the  model), the  particle
number in the NESS must be  infinite. However, most of the particles will have
very small kinetic  energy, so that the system has  very nice energy profiles,
which, so  to speak, are  generated by those  particles with energy  away from
0. This  is in  fact  reminiscent of  non-normalizable  measures in  dynamical
systems \cite{collet90}.

The model  in question consists of a  chain of identical cells,  each of which
contains  a fixed  point-like scatterer  that exchanges  momentum  with tracer
particles.   Inside the  system the  particles move  deterministically between
cells, interacting with  the scatterers but not among  themselves. However, on
their passage the  particles modify the local state  of the substrate, which
in  turns alters  the evolution  of the  other particles.   At  the collisions
energy  is conserved.   At its  boundaries the  chain is  in contact  with two
stochastic  particle  reservoirs, characterized  by  a  fixed temperature.

In the next section  we describe
in detail  the model and in \sref{sec:noneq} the general properties of its 
non-equilibrium steady state. In \sref{sec:comp} we study the continuous limit 
of our model and compare it with the theory that appeared in \cite{CE08}. Finally, in 
Sects.~\ref{sec:E-transport} and \ref{sec:kappa} we discuss the energy and particle transport of the 
deterministic finite chain.

\section{The Model}
\label{sec:model}

In this section, we describe the 1-dimensional particle model that we
consider, and briefly review the results of \cite{CE08} for the continuous
model.

The model consists  of $N$ cells in  a row, each cell of  length $\lambda$. In
the center  of each cell there is  a point-like scatterer which  does not move
but which has a ``momentum'' $P\in  \mathbb{R}$ and a mass $M$. Particles move
in these  cells.  They have mass $m\ne  M$ and momentum $p$.  The particles do
not  interact among themselves  but they  do interact  with the  scatterers as
follows:  Whenever a  particle with  momentum  $p$ reaches  a scatterer  whose
momentum  is $P$, the  following happens:  With probability  $1-\gamma/N$, the
particle  crosses to  the  other side  of  the scatterer,  and continues  with
momentum $p$, while the scatterer  retains its momentum $P$.  With probability
$\gamma/N$ actual  scattering takes place and  the new momenta  $\tilde p$ and
$\tilde P$ are given by
\begin{equ}
\begin{pmatrix}
\tilde p \\ \tilde P\\
\end{pmatrix}
=
S
\begin{pmatrix} 
p\\ P
\end{pmatrix}~,
\end{equ}
where the scattering matrix $S$ is
\begin{equ} \label{eq:col-rules}
S=
\begin{pmatrix}
-\sigma &1-\sigma\\ 
1+\sigma& \sigma
\end{pmatrix}
~,\quad \text{and}\quad\sigma=
(M-m)/(M+m)~.
\end{equ}
When $\gamma=N$,  the particles  interact with the  scatterer every  time they
encounter one, and the model is  fully deterministic, except for the nature of
the  baths. These rules  are similar  in spirit  (but far  more rich),  to the
flipping  Lorentz lattice gases  studied some  years ago  \cite{boon00}.  When
$\gamma<N$,  then the  model  has some  randomness,  since we  need to  decide
whether scattering takes  place, or the particle flies  through the scatterer.
Note, however, that this randomness does not change the energies of the actors
in the system.

The  collision  rules are  just  those of  elastic  scattering, but
with the scatterers not moving.  The deterministic  model (with
$\gamma=N$)  is a  one  dimensional generalization  of  the models  previously
studied  in \cite{MMLL01,EY06,LLMM03,EMMZ06}, where  the scatterers  are fixed
freely rotating disks. In these  models, the collisions provide a local energy
mixing among different degrees of freedom that leads to a local state which is
well approximated by a local equilibrium state where the particles behave as a
perfect  gas.  Close to equilibrium,  Green-Kubo relations  for  the heat  and
particle fluxes are valid  and the corresponding Onsager reciprocity relations
are  satisfied  \cite{MMLL01,LLMM03}. Moreover,  in  the zero-coupling  limit,
where  the  invariant measure  of  the NESS  is  expected  to be  multivariate
Gaussian,  compact analytical  expressions for  the currents  and  the density
profiles have been obtained \cite{EY06}.  To the lowest order, the corrections
due  to a  finite coupling  were considered  in \cite{EMMZ06}.   Note  that if
particles and scatterers have the same  mass, the model has an infinite number
of  conserved  quantities,  as  their   momenta  are  just  exchanged  in  the
collisions.   Trajectories for  one single  particle have  been  considered in
\cite{BK03}.

To force the  system out of equilibrium, we couple  the leftmost and rightmost
cells to infinite ideal particle reservoirs. From each reservoir particles are
injected  into the  system  at a  given  rate $\nu$  and with momenta  distributed
according to
\begin{equ} \label{eq:MB}
F(p) \ dp \ = \ \Theta(\pm p)\left(2\pi m\kB T\right)^{-1/2} e^{-p^2/2m\kB T}
\ dp\ , 
\end{equ}
where $T$  is the temperature of  the reservoir, $\kB$  the Boltzmann constant
and  the Heaviside  function $\Theta(p)$  restricts the  sign of  the momentum
according to the side from where the particles are injected (``$-$'' for those
entering  from  the  right  side  and  ``$+$'' for  those  entering  from  the
left). Equation \eref{eq:MB} implies that the particles in the vicinity of the
opening between  the system and the  reservoir have a  momentum density $f(p)$
with a non-normalizable singularity at $p=0$. For the reservoir coupled to the
leftmost ($i=1$) cell, $f(p)$ is
\begin{equ} \label{eq:fL}
f_\L(p) \ dp \ = \ \frac{\Theta(+p)}{\left(2\pi m \kB
 T_\L\right)^{1/2}} \frac{e^{-p^2/2m\kB T_\L}}{|p|} \ dp\ ,
\end{equ}
and for the reservoir coupled to the rightmost ($i=N$) cell
\begin{equ} \label{eq:fR}
f_\R(p) \ dp \ = \ \frac{\Theta(-p)}{\left(2\pi m \kB T_\R\right)^{1/2}}
\frac{e^{-p^2/2m\kB T_\R}}{|p|} \ dp\ . 
\end{equ}

The reader should  note that the singularity of  $f(p)$ is non-integrable only
for one  dimensional models, more  precisely, for models with  one dimensional
dynamics.  This is  related to  the fact  that, while  in any  dimension, slow
particles  need  much  longer  times  to  move across  the  system  than  fast
particles,  only in  1-dimensional  dynamics does  this  imply that  particles
steadily accumulate  near $p=0$. In higher dimensions,  $1/|p|$ is integrable
near $p=0$.  Thus, the particle density  diverges in the  stationary state as
$t\to\infty$. Due to  this seemingly unphysical property, in  the past, it has
been  argued   that  injecting   particles  with  the   momentum  distribution
\eref{eq:MB} cannot be correct \cite{koplik98}.  However, as has been shown in
\cite{CE08},  it is  the reservoir  distribution  $f(p)$ and  not $F(p)$  that
admits stationary solutions which, at the same time, preserve the distribution
of  the  scatterers' momenta.  Furthermore,  the  higher  momenta of  $p$  are
obviously well-behaved in the limit of $t\to\infty$.

\begin{figure}[!t]
\begin{center}
 \includegraphics[scale=1]{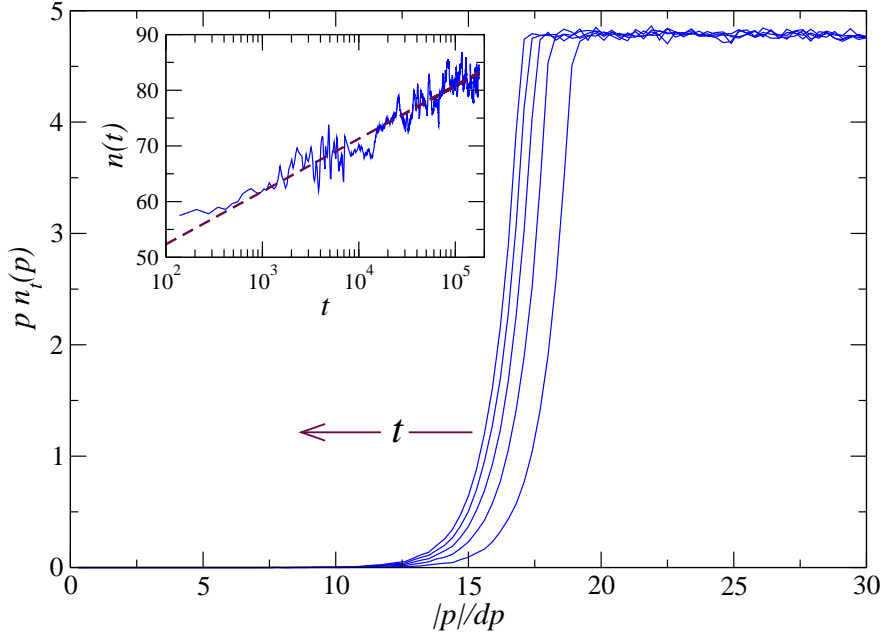}
\caption{Number  of particles  $n_t(p)$ with  momentum $\approx  p$  times the
momentum $p$,  at $t=200$, $600$,  $1000$, $1400$ and  $1800$, for a  chain of
$N=201$  cells at  equilibrium with  $\nu_\L=\nu_\R=100$  and $T_\L=T_\R=100$,
$\sigma=1/2$ and $\gamma=N$. $dp$ is the width of the bins used to compute the
empirical  distribution $n_t(p)$. Thus,  the $x$-axis  corresponds to  the bin
number. In the inset: logarithmic  divergence of the total number of particles
inside the system.
 \label{fig:n-vs-p}}
\end{center}
\end{figure}

More precisely, denoting  by $g(P_i)$ the distribution of  the momentum of the
$i$-th   scatterer,     it  was   shown  in  \cite{CE08} that   at
equilibrium
($T_\L=T_\R=T$), both $F(p)$ and $g$ are Gaussian and are given by
\begin{equ} \label{eq:f-eq}
F(p) \ dp \ = \ {\left(2\pi m \kB T\right)^{-1/2}}
{e^{-p^2/2m\kB T}} \ dp\ , 
\end{equ}
\begin{equ} \label{eq:g-eq}
g(P) \ dP \ = \ \left(2\pi M\kB T\right)^{-1/2} e^{-P^2/2M\kB T} \ dP\
. 
\end{equ}
As mentioned before, because of the factor $1/|p|$ in $f$, the number of particles in the
system is infinite at stationarity. Starting with any initial
distribution, and letting the system evolve in time $t$, the number of
particles with low momenta grows without bounds. On the other hand
if one can prove that  the number of particles
with momentum $p>p_0$  has a limit as $t\to\infty$  then the stationary state
is well defined,  in spite of the divergence of the  total number of particles
\cite{collet90}. To convince  ourselves that this is indeed  the case, we have
measured the  evolution in time of the  number of particles in  the system and
their  momentum distribution,  for a  chain of  $N=201$ cells  at equilibrium:
particles were injected  from both reservoirs at the  same rate $\nu=100$ with
the same temperature $T=100$.  The results are shown in \fref{fig:n-vs-p},
where the  number of  particles $n_t(p)$ with  momentum $\approx p$  times the
momentum $|p|$, is  plotted for several successive times.  We observe that for
$p$ larger  than some  $p_0(t)$, $p n_t(p)$ is  stationary. This means  that the
total number  of particles  $n(t)=\int_{-\infty}^\infty n_t(p) \  dp$ diverges
due  to the  slow  accumulation  of cold  particles.  Indeed, $n(t)$  diverges
logarithmically   with   time,   as   can    be   seen   in   the   inset   of
\fref{fig:n-vs-p}. In the same vein,  $p_0(t)$ is seen to decrease   to  zero
logarithmically.

In the limit of $N\to\infty$,  setting $x=i/(N\cdot \lambda)$ for the position
of the  $i$-th cell, $g(P_i)\to g(P,x)$,  is the probability  density that the
scatterer at  position $x$ has  momentum $P$ and  so $\int dP  \,g(P,x)=1$, by
definition. In  \cite{CE08} it was argued  that in the  continuous limit, this
system can  be modeled by a  Boltzmann equation, whose  stationary solution is
described by the equations \minilab{boltzmann}
\begin{equs}
p\partial_x F(p,x) =&\,\, \gamma |p| \int {dP}\,
\left(F(\tilde p,x)g(\tilde P\,,x)-F(p,x)g(P\,,x)\right)~,\label{ba}\\
0 = & \int {dp}\,
\left(F(\tilde p,x)g(\tilde P\,,x)-F(p,x)g(P\,,x)\right)~,\label{bb}\\
\end{equs}
where the quantity $F(p,x)$ is equal to $F(p,x)=|p| f(p,x)$, with $f(p,x)$ the
probability  density that a  particle at  position $x$  has momentum  $p$, and
$\gamma\in[0,N]$ corresponds to the same quantity of the particle model, which
determines  the  scattering probability.  Therefore,  for  $p>0$, $F(p,x)$  is
related to  the rate  of particles with  momentum $p$  moving from $x$  to the
right. Similarly,  $F(p,x)$ is,  for $p<0$, related  to the rate  of particles
with  momentum $p$ moving  from $x$  to the  left. Note  that, in  contrast to
$f(p)$, the function  $F(p)$ is free of singularities.  Therefore, in spite of
the infinite  number of  particles in the  stationary state of  the scattering
model, the flux $F(p,x)$ is finite and integrable, and it is for this quantity
that the Boltzmann equation is formulated.

The Boltzmann  equation \eref{boltzmann} was derived assuming  that $F(p)$ and
$g(P)$ are  statistically independent.  Therefore, the  similarity between the
particle model and its Boltzmann version should be best in the case of large $N$ and
when  there are  many  particles  (with momentum  $|p|>p_0>0$)  in each  cell.
Moreover,  in \cite{CE08},  it was  proven  that for  any particle  injections
$f_\L(p)$ and  $f_\R(p)$ in a  certain cone in Banach  space, \eref{boltzmann}
has solutions  when $\gamma=\OO(1)$.  Our  numerical studies are,  however, for
parameter values well outside this cone, and still give a very good comparison
between  the  particle  model  and  the Boltzmann  model.   Furthermore,  when
$\gamma=\OO(N)$, the particle model  is still well defined, although different
from the  Boltzmann model. If $\gamma=N$  a particle will  scatter whenever it
meets a scatterer  and will never fly to the other side of a  scatterer without collision. When  $\gamma \ll N$, the local state cannot  correspond to a local
equilibrium state.   Indeed, when scattering is  rare, the particles do  not interact,
thus leading  to local states that are  described by the sum  of two different
families  of  particles:  those  that  were injected  from  the  left,  flying
ballistically  to  the right,  and  those injected  from  the  right that  fly
ballistically to  the left \cite{DD99}.  However, if  $\gamma=N$, then all particles
scatter  when  they   encounter  a  scatterer,  leading  to   a  stronger  local
interaction. One  would expect that this  strong interaction would  lead to a
diffusive  particle behavior. However,  in  \sref{sec:E-transport}, we
will show that the transport remains superdiffusive.

\section{Non-equilibrium steady state}
\label{sec:noneq}

In this section we consider the model described
in the  previous section, coupled  to reservoirs injecting particles  into the
chain, at different rates and at different temperatures.
In the rest of the paper we will study our model
for $\gamma=1$ (corresponding to the Grad limit considered in \cite{CE08}) and 
for $\gamma=N$ (corresponding to the deterministic particle
dynamics).

Out of equilibrium, particle and  energy currents appear, whose magnitudes are
determined  by the  differences of  injection rates  and of  temperatures. The
particle injection rate is defined as
\begin{equ} \label{eq:nu}
\nu =\int_0^\infty F(p) dp \ ,
\end{equ}
where  $F(p)$  is the
momentum distribution of the injected  particles, given by \eref{eq:MB}. 
As we
have discussed, the particle density  in the bath is  infinite. However, 
since  the integral in  \eref{eq:nu} is
finite, one realizes that $\nu$  takes  the  role  of  an  effective  
particle  density.
Moreover,  since  the  injected  particles  correspond to  a  perfect  
gas  at
equilibrium, one can define the chemical potential of the reservoir as
\begin{equ} \label{eq:chempot}
\mu = \mu_0 + T\log\left(\frac{\nu}{T}\right) \ ,
\end{equ}
with  $\mu_0$  a  pure  constant.  The injection  rate  $\nu$  also  trivially
determines the rate at which energy is injected into the system as
\begin{equ} \label{eq:eps}
\varepsilon = \nu \kB T \ .
\end{equ}

\begin{figure}[!t]
\begin{center}
 \includegraphics[scale=1]{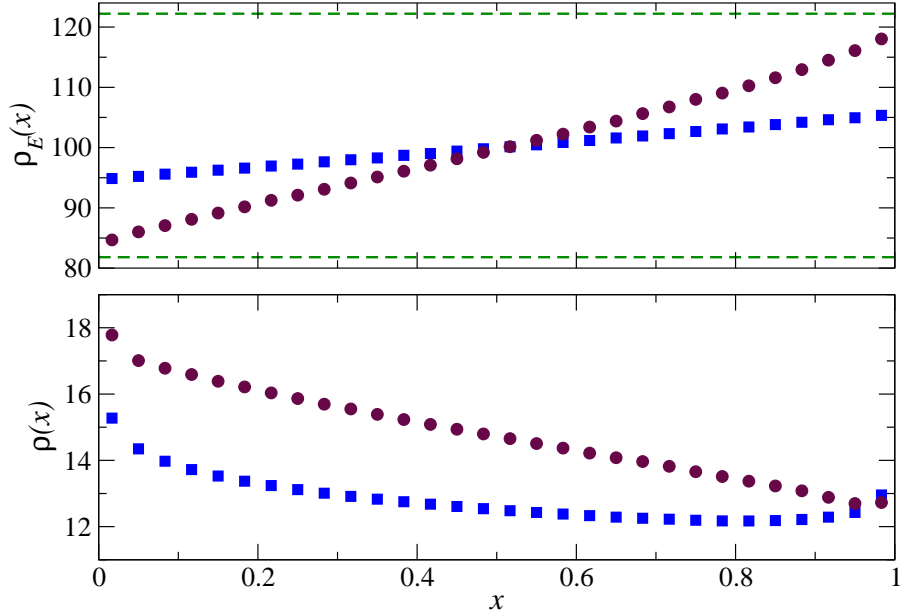}
\caption{Profiles of the particle density (lower panel) and the kinetic energy
 per  particle density  $E_K$  (upper panel),  for  a chain  of $N=30$  cells,
 $\sigma=0.5$  and   for  $\gamma=1$  (squares),   and  $\gamma=N$  (circles).
 Particles were  injected to the chain with  rates $\nu_\L=220$, $\nu_\R=180$
 and  temperatures $T_\L=81.818$  and  $T_\R=122.222$, indicated  here by  the
 dashed   line.   Here,   $T(x)$   is   computed  as   the   mean   scatterers
 kinetic. \label{fig:profs}}
\end{center}
\end{figure}

We  now  proceed   to  study  the  non-equilibrium  state   of  our  model.  In
\fref{fig:profs}  we   compare  the   particle  density  $\rho(x)$   and  the
temperature  profiles numerically  obtained for  two different  values  of the
scattering parameter, $\gamma=1$ and $\gamma=N$. The temperature at the $i$-th
cell is  computed as the average  $p^2$ with respect to  the particle momentum
distribution function  $F_i(p)$ measured at the $i$-th  cell. This temperature
coincides  with the  time averaged  kinetic  energy of  the $i$-th  scatterer,
indicating a good local equilibration. All the observables were averaged for a
time interval  during which the total  number of particles  inside the channel
does not change appreciably.

The  temperature  in  the bulk  of  the  system  does  not match  the  nominal
temperatures of  the reservoirs  (indicated by the  dashed lines in  the upper
panel). Moreover, we observe that the  energy mismatch at the contact with the
reservoirs  depends   on  the  scattering  probability.   For  $\gamma=1$  the
temperature profile  is less steep than  for $\gamma=N$ as  expected, since in
the former case, particle-particle interaction, mediated by the scatterers, is
less effective than in the latter case. As for the particle density, a smaller
scattering probability  leads to a  less steep density profile.  Moreover, the
accumulation  of  particles  at  the  contact  with  the  reservoirs  is  more
pronounced  when  the  scattering  is  less  frequent.  This  is  because  the
scattering contributes  to heat up  the injected cold  particles. In the  limit of
$N\to\infty$,  the accumulation  of particles  at the  boundaries produces
singularities \cite{koplik98}.

We  also have measured  the rate  at which  particles cross  from one  cell to
another from left to right $\nu_\R(x)$ and from right to left $\nu_\L(x)$.

Out of equilibrium,  the particles inside the cell at  position $x$, leave the
cell to the  right at a different rate  than the rate at which  they leave the
cell  to  the   left.   This  is  a  consequence   of  the  substrate-mediated
particle-particle interaction. Calling the  rate at which particles cross from
one cell  to another  from left to  right $\nu_\R(x)$  and from right  to left
$\nu_\L(x)$, the particle current is defined as
\begin{equ} \label{eq:J}
J_n(x) = \nu_\R(x) - \nu_\L(x+dx) \ .
\end{equ}
Therefore, these local rates are strongly dynamically constrained, so that the
stationary  particle  current  is  uniform.   Indeed, we  find  that,  in  the
stationary state,  a uniform particle current, with  extremely linear profiles
for $\nu_\R(x)$ and $\nu_\L(x)$. In \sref{sec:E-transport}, we will see
that this strong correlations  determine an unexpected superdiffusive particle
and energy transport.

\section{Infinite volume limit: $\gamma=1$}
\label{sec:comp}

In  this  section,  we  compare the  non-equilibrium  probability  distribution
functions  of  the discrete  model with  those  predicted by  the  Boltzmann
equation \eref{boltzmann}.

We have  solved  numerically \eref{boltzmann}  by a discretization  in momentum
space. Fixing
the spacing  of the discretized momentum  to $\Delta p$, the  number of points
we considered is the  minimum necessary  to keep  the information  from the
tails of the distributions as small as $\sim10^{-10}$. We proceed as follows: at
any  $x$, the equation  \eref{bb} for  $g$ only depends on
$F(\cdot,x)$. We discretize \eref{bb} and solve it
as an eigenvalue  problem (with eigenvector $g(\cdot,x)$).  The only
limitation is the size of the matrices one obtains in this way: Our
runs were done with matrices of size $\sim 4200$. The function
$g$ found in this way  is then inserted into \eref{ba}. High-order integration
in position  space is then  used to integrate  \eref{ba} from $x=0$  to $x=1$.
Therefore, fixing  the injection  at $x=0$ to  \eref{eq:fL}, we use  a shooting
method to determine the extraction of particles at $x=0$ in such a way that at
$x=1$ the desired injection \eref{eq:fR}  will result (see also \cite{CE08} for
more details).

\begin{figure}[!t]
\begin{center}
 \includegraphics[scale=1]{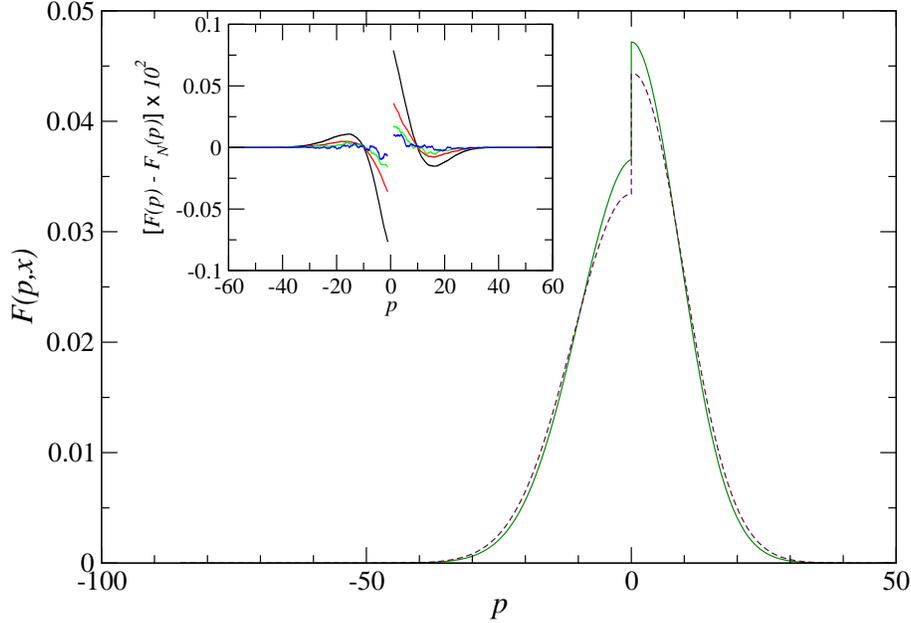}
\caption{Momentum distribution  $F(p,x)$ of the  particles at the  left (solid
  curve) and  right (dashed curve) ends of  the system , from  the solution of
  the  Boltzmann equation  \eref{boltzmann},  for $\sigma=0.5$,  $\nu_\L=220$,
  $\nu_\R=180$,  $T_\L=81.818$ and  $T_\R=122.222$.  In  the inset  the finite
  size deviation of $F_N(p,x)$ from the Boltzmann solution are shown for $N=2$
  (black), $N=4$ (red), $N=8$ (green) and $N=16$ (blue).
\label{fig:Pv-dev}}
\end{center}
\end{figure}

In  \fref{fig:Pv-dev}  we  show   the  solution  of  \eref{boltzmann}  for
$\sigma=0.5$, $\nu_\L=220$, $\nu_\R=180$, $T_\L=81.818$ and
$T_\R=122.222$, and $\gamma=1$.   The  first
peculiarity of the non-equilibrium distributions  is the jump at $p=0$. This is
due   to   the   very   weak  particle-particle   interaction   obtained   for
$\gamma=1$. The size  of the jump is partly determined by  $\gamma$ and, as it
is clear from Eq.~\eref{eq:nu}, partly  by the difference between $\nu_\L$ and
$\nu_\R$.  Note  that, as  a  consequence  of  the temperature  gradient,  the
positive  and  negative  parts  of  the distribution  are  only  approximately
Gaussian. They are Gaussian only if $T_\L=T_\R$.

To  study  the limit  $N\rightarrow\infty$,  we  have  numerically
followed the  evolution of  finite size $N$  chains and measured  the particle
momentum  distribution $F_N(p,x)$  for the  same parameters  as above.  In the
inset of \fref{fig:Pv-dev}, the deviation of $F_N(p,x)$ from the solution
of the Boltzmann equation $F(p,x)$ is shown for chains from $N=2$ to $N=16$.

\begin{figure}[!t]
\begin{center}
 \includegraphics[scale=1]{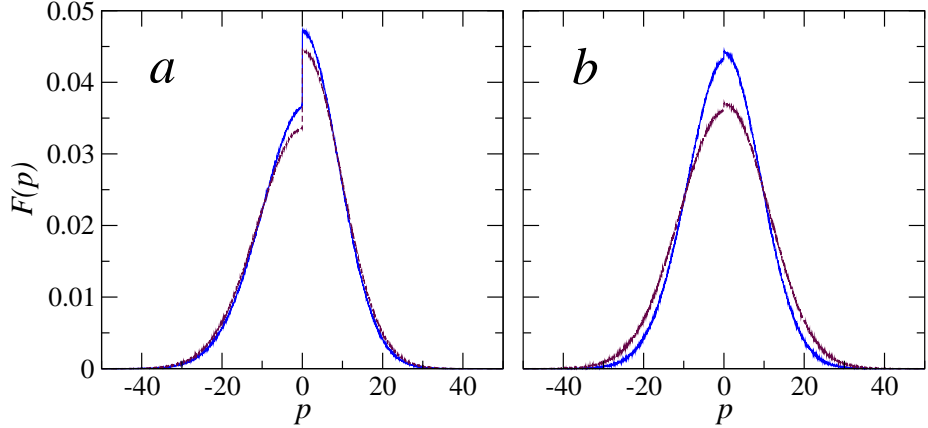}
\caption{Momentum distribution  $F_N(p,x)$ of the particles  at $x=1/N$ (solid
 curve)  and  $x=1$  (dashed  curve),  for  a  chain  of  $N=30$  cells,  with
 $\sigma=0.5$, for $a)$ $\gamma=1$  and $b)$ $\gamma=N$. The bath's parameters
 are as in \fref{fig:Pv-dev}.
\label{fig:Pv-inj}}
\end{center}
\end{figure}

Deviations are  seen over the whole  domain, although they are  biggest at the
center  of the  distribution.  They  are  not symmetric  in $p$,  which is  an
indication that, for a given size  $N$, deviations may depend on the injection
rates and  bath's temperatures in  general.  Furthermore, we observe  that
the  solution of  \eref{boltzmann} appears  to be  the asymptotic
distribution, $\lim_{N\rightarrow\infty} F_N(p,x) \rightarrow F(p,x)$.  In any
case, the deviations from $F(p,x)$ are less than $0.1\%$, tending to zero very
fast.  For instance,  for  a chain  of  $N=16$ the  deviations  are less  than
$0.01\%$.

Finally,  we   have  also  measured   the  distribution  $F_N(p,x)$   for  the
deterministic  finite chain  ($\gamma=N$). In  \fref{fig:Pv-inj},  we show
$F_N(p,x)$ at $x=1/N$  (solid curve) and $x=1$ (dashed curve),  for a chain of
$N=30$ cells and  $\gamma=1$ (panel $a$), $\gamma=N$ (panel  $b$). The
other parameters are reported in the caption  of \fref{fig:Pv-dev}.  The
distributions  $F_{30}(p,x)$ in  \fref{fig:Pv-inj}-$a$,  are on  top of  the
solution  \eref{boltzmann}  and, as  mentioned  above,  the  deviations from  the
asymptotic distribution decay very fast.  For $\gamma=N$, the jump at $p=0$ is
much  smaller, with the only  remaining the  contribution  coming from
the  difference $|\nu_\L - \nu_\R|$ of  the
injection rates.   As expected, the distributions in
\fref{fig:Pv-inj}-$b$ are not Gaussian for this non-equilibrium case.

\section{Energy transport}
\label{sec:E-transport}

In  this  section  we  turn  our  attention  to  the  heat  transport  of  the
deterministic  model,  {\it i.e.},   setting  $\gamma=N$.  
The particles of mass $m$ (which we call tracers) are the only energy
carriers of the system. We start by analyzing their dynamics.

\subsection{Microscopic evolution}
\label{sec:micro}

We consider a finite chain of  $N$ cells with periodic boundary conditions and
$n$  particles {\emph{per  cell}}. A  stationary  state of  this closed  system is  the
equilibrium  state  characterized  by  $N$,  $n$ and  the  total  energy  $E$
given by
\begin{equ} \label{eq:total-E}
E_0 = \frac{1}{2}\left(\frac{1}{m}\sum_{i=1}^n p_i^2 + \frac{1}{M}\sum_{i=1}^N
P_i^2\right) \ . 
\end{equ}
We start the evolution with the scatterers at 0
momenta. As for 
the open  chain, the state of the system approaches  equilibrium 
logarithmically  slowly (in time). 
During  the  transient, the
substrate continuously  extracts energy from  the gas of particles;  the total
energy of the scatterers grows  logarithmically in time, until it saturates at
sufficiently  long times.  All measurements  are  taken after  the system  has
relaxed to the approximate equilibrium state.

Once equilibrium is reached, we focus on the evolution of
a  tagged  particle in  the  bulk  of  the system,
when $N$ and $n$ are sufficiently large. This is convenient,  since
particles  interact  among  themselves  only through  their
collisions  with  the 
substrate, and thus the  local dynamics depends on  the particle
density.   Here, we are 
interested in  the high density regime, which,  following the discussion
about  \fref{fig:n-vs-p},  is  a  good  approximation  of  the  stationary
$n=\infty$ state. 

\begin{figure}[!t]
\begin{center}
 \includegraphics[scale=1]{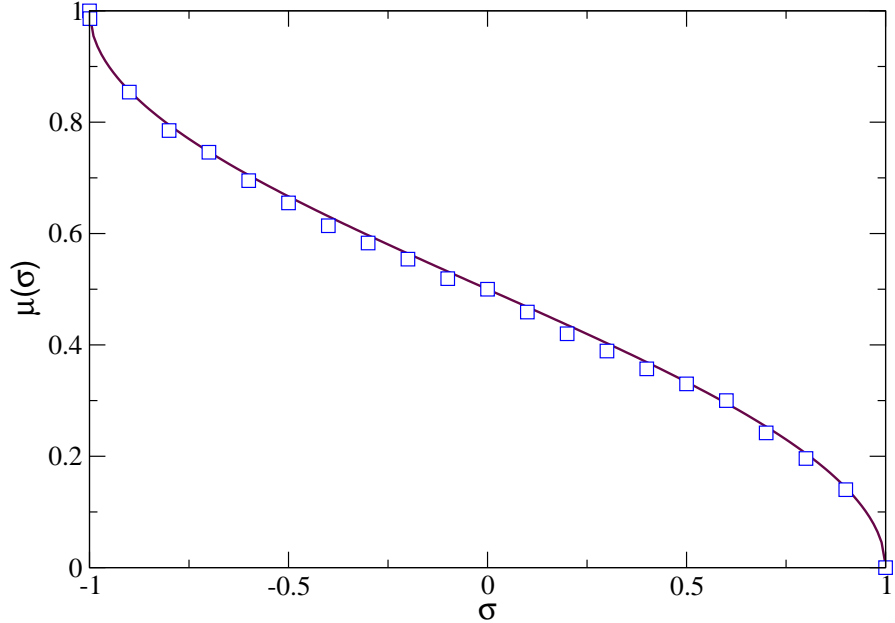}
\caption{Persistent  probability  $\mu$  as  a  function  of  the  mass  ratio
  parameter $\sigma$,  averaged over the evolution  of a tagged  particle in a
  chain of $N=11$, $n=20$ and $T=500$.
\label{fig:persistent}}
\end{center}
\end{figure}

When the particle  encounters a scatterer, its velocity after
collision is
determined by \eref{eq:col-rules}.  In fact,
this scattering  matrix leads to a  persistent motion of  the particle, namely
the probability that the particle continues in the same
direction in which it reached the scatterer, is  larger than $1/2$.
This probability, $\mu$,  can be 
easily computed as follows\footnote{To analyze  the dependence of $\mu$ on the
masses,  it is  convenient to  work with  the velocities  instead of  the
momenta.}: without  loss of generality  assume that before the  collision, the
particle's  velocity  is  $v>0$.    In  equilibrium,  the  scatterer  velocity
distribution function is obtained  by taking $V=P/M$ in \eref{eq:g-eq}. Taking
into account that after the collision the particle's velocity is $v' = -\sigma
v + (1+\sigma) V$ (see \eref{eq:col-rules}), the probability that after the
collision the particle has a velocity $v'>0$ can be written
as\footnote{There is no factor $1/|v|$ here, because we must consider
the probability of a particle with velocity in $[v,v+dv]$ hitting a
scatterer within a given time.}
\begin{equ} \label{eq:pers-1}
\mu(\sigma) \equiv P(v'>0 | v>0) = \frac{(m M)^{1/2}}{2\pi\kB T}\int_0^\infty dv~~
e^{-\frac{mv^2}{2\kB T}}\int_{\frac{\sigma}{1+\sigma}v}^\infty dV
e^{-\frac{MV^2}{2\kB T}} \ ,
\end{equ}
that can be integrated to yield
\begin{equ} \label{eq:pers-2}
\mu(\sigma) = \frac{1}{2} - \left(\frac{m}{2\pi\kB
  T}\right)^{1/2}\int_0^\infty dv~\mathrm{erf}\left(\frac{(M-m)v}{(8M\kB
  T)^{1/2}}\right) e^{-\frac{mv^2}{2\kB T}} \ ,
\end{equ}
where $\mathrm{erf}(\cdot)$ is  the error function. The limit  values of $\mu$
can be easily  read from \eref{eq:pers-2}: for $\sigma=-1$,  namely $M=0$, the
error  function is  $\mathrm{erf}(-\infty)=-1$ and  $\mu=1$.  In  the opposite
case, when  $\sigma=1$ ($M=\infty$), $\mu=0$. Finally,  for $\sigma=0$, namely
$M=m$, $\mathrm{erf}(0)=0$ and $\mu =  1/2$. With the exception of $\sigma=0$,
the dynamics of the particles is persistent. In \fref{fig:persistent}, the
probability $\mu(\sigma)$,  computed from the statistics of  the collisions of
the tagged particle is shown.

\begin{figure}[!t]
\begin{center}
 \includegraphics[scale=1]{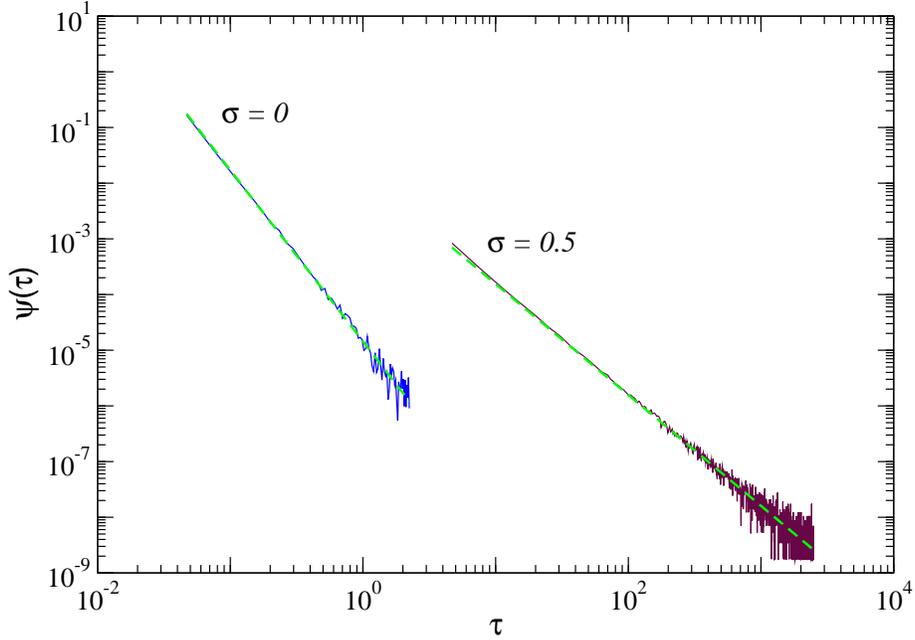}
\caption{Distribution  function  of  the  collision  times  $\Psi(\tau)$,  for
  $\sigma=0$ and $\sigma=0.5$.  The dashed lines correspond to  fit to a power
  law. We obtain, in  within numerical accuracy, $\Psi(\tau)\sim\tau^{-3}$ for
  $\sigma=0$ and $\Psi(\tau)\sim\tau^{-2}$ for $\sigma=0.5$.
\label{fig:levy}}
\end{center}
\end{figure}

At a coarse grained description, the dynamics of the particles can be seen as a
persistent random walk with waiting time $\tau$ corresponding to the collision
times, that  are determined by the  particle's velocity. We  have measured the
distribution  of the  waiting  time  $\Psi(\tau)$ of  the  tagged particle  for
different values of $\sigma$.  In \fref{fig:levy} we show $\Psi(\tau)$ for
$\sigma=0$  and $\sigma=0.5$.   As a  consequence of  the  single particle's
velocity  distribution, $\Psi(\tau)$  turns  out to  be  a broad  distribution
$\Psi(\tau)\simeq   \tau^{-(1+s)}$,  with   $s\simeq1$  for   $\sigma=0.5$  and
$s\simeq2$ for $\sigma=0$. Therefore, our persistent walker seemingly performs
a Levy walk. The power $-2$ for $\sigma\ne 0$ can be derived
(approximately) from a
multiple integral as in \eref{eq:pers-2}.

\begin{figure}[!t]
\begin{center}
 \includegraphics[scale=1]{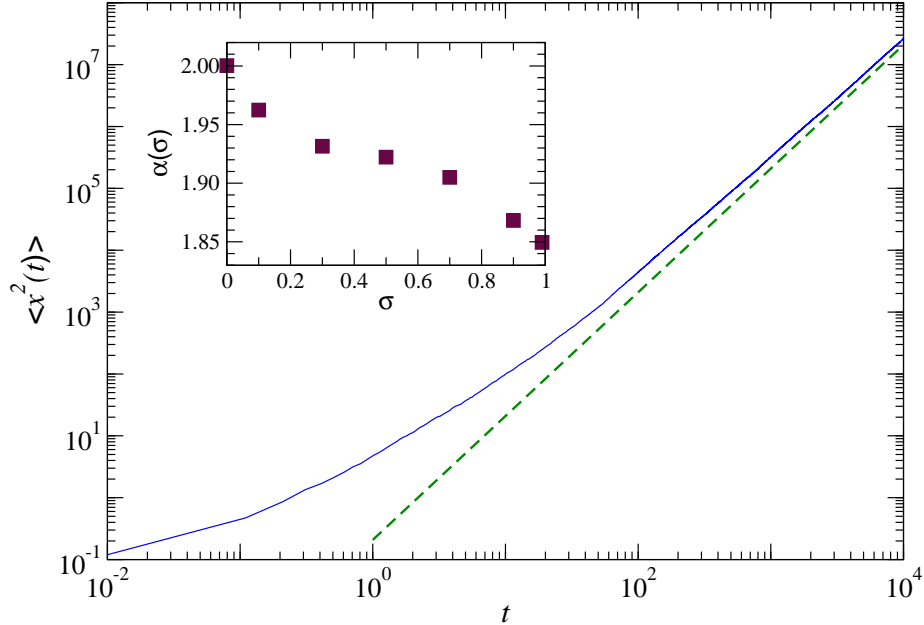}
\caption{Time  dependence  of the  variance  of  the  position of  the  tagged
particle  $\langle  x^2(t)\rangle$,  for  $N=11$, $n=20$  and
temperature $T=500$  (solid
curve), when $\sigma=1$. The  dashed line  corresponds to  a scaling $\sim  t^2$. In  the inset,
asymptotic power scaling $\langle  x^2(t)\rangle \sim t^\alpha$, as a function
of the mass ratio parameter $\sigma$.
\label{fig:var}}
\end{center}
\end{figure}

In  the continuous limit,  a persistent  random walker  yields to  a particle's
density   whose  evolution   is  described   by  the   telegraph  equation
\cite{weiss}. Noting that asymptotically  the telegraph equation yields to
a diffusive  evolution and  that the  Levy walk for  $1<s<2$ does  only induce
anomalous corrections  to the normal long-time  behavior \cite{bouchaud}, one
would  expect  that  the  microscopic  particle's  dynamics  yields  diffusive
transport. However,  this is not the  case. In \fref{fig:var}  we show the
evolution  of the dispersion  of the  position of  a tagged  particle $\langle
x^2(t)  \rangle$  for  $\sigma=0.5$,  averaged  over an  ensemble  of  initial
conditions.   Asymptotically,  $\langle x^2(t)  \rangle  \sim t^\alpha$,  with
$\alpha \lesssim  2$. In fact,  we have found  that the asymptotic  scaling of
$\langle x^2(t)  \rangle$ depends  on the mass  ratio parameter  $\sigma$ (see
inset  of  \fref{fig:var}).   For  $\sigma=0$  the  particle's  motion  is
ballistic, while for  $\sigma\ne 0$,  the motion is
superdiffusive. The observed anomalous behavior proves that in one dimension,
the effect of the dynamical memory of the deterministic model is much stronger
than in  higher dimensions\footnote{As a note  aside, if the  direction of the
particle after  the collision if  chosen randomly so  that the effects  of the
dynamical  memory   can  be  neglected,   then  the  diffusive   transport  is
recovered.}. Since the particles are the energy carriers one expects that
the energy transport will be anomalous as well. We study this in the
next section.

\section{Heat conductivity}
\label{sec:kappa}

\begin{figure}[!t]
\begin{center}
 \includegraphics[scale=1]{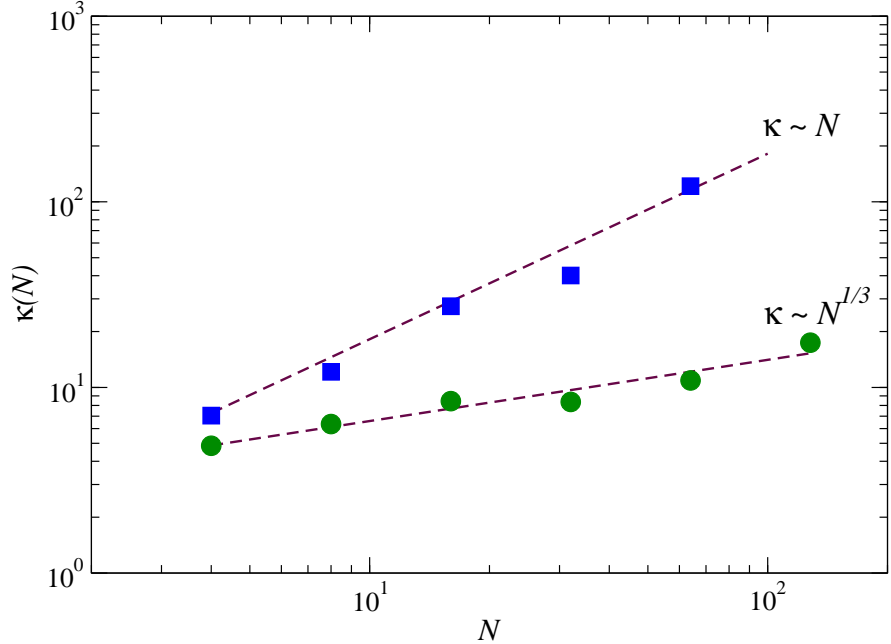}
\caption{Heat conductivity $\kappa$, as a function of the number of cells $N$,
 for two values of $\gamma$: $1$  (squares) and $N$ (circles). The rest of the
 system's parameters are as in \fref{fig:Pv-dev}. For $\gamma=N$, $\kappa$
 diverges as $\sim N^{1/3}$, while  for $\gamma=1$, $\kappa$ diverges as $\sim
 N$. These power laws are indicated by the dashed lines. \label{fig:kappa}}
\end{center}
\end{figure}

We turn  our attention to the  energy transport of our  model. Considering the
open  system coupled at  its boundaries  to two  particle reservoirs,  we have
computed the dependence  of the heat conductivity $\kappa$ on  the size of the
system $N$, for fixed nominal values of the injections and temperatures of the
particle reservoirs.  We define the heat conductivity as
\begin{equ}
\kappa = \frac{J_U}{T_N - T_1} \ ,
\end{equ}
where $J_U$  is the  measured energy  current and $T_1$  (resp.~$T_N$)  is the
temperature measured in  the leftmost (resp.~rightmost) cell  that, as we have
seen, in  general does not coincide  with the temperatures  of the
reservoirs. (The length of the system is 1 since we space the
scatterers by $\lambda =1/N$.)
The  results of simulations are shown  in \fref{fig:kappa}  for $\gamma=1$  (squares) and
$\gamma=N$ (circles). When the scattering is rare ($\gamma=1$), we obtain
$\kappa\sim N$.   This can be understood from  the fact that particles  move ballistically,
interacting with the lattice very rarely.

Surprisingly, the energy transport  in the deterministic model ($\gamma=N$) is
anomalous, with a heat conductivity that diverges as $\kappa \sim N^{1/3}$. It is
interesting to note that usually anomalous heat conduction is related to the existence
of additional global conserved  quantities \cite{LLP03}.  However, for $\sigma
\ne  0$ the bulk  dynamics of  our model  only  conserves
energy. As far as we know, this is the first example  of a mechanical model
that,  not  having  additional  integrals  of  motion,  shows  anomalous  heat
conduction.

\subsection{Return to equilibrium}
\label{sec:return-equil}

In order to shed more light on the  anomalous heat  transport we  studied  the system's
equilibrium  response to  a finite  energy  perturbation.  Suppose  that at  a
certain initial  time, $t=0$, the equilibrium state of  the system is
perturbed by  an additional  amount of energy  $\Delta E$ that  is distributed
among all the degrees of freedom in  a finite region of volume $V$, around the
position $x$.  By measuring
the  evolution  of the  energy  field, one  can  estimate  how heat
propagates through the system.

Considering  the closed  system as  in \sref{sec:micro} we  proceed as
follows: at time $t=0$ we perturb the  state $S^0$ of the system to $\tilde{S}$,  as follows:  the  energies  of  the particles  and  scatterers
contained in  the $\mathcal{N}$  central cells are changed so that  the total
energy  inside these  cells  is $E_{\rm pert}$.  After  this, we  let the  central
subsystem relax. 
To obtain the evolution  of the energy perturbation $\Delta E(x,t)$, we
have followed  two trajectories of the system: the unperturbed one,  with initial
state $S^0$ and  the perturbed one, with initial  state $\tilde{S}$. Then, the
energy difference at time $t$ and position $x$ is
\begin{equ} \label{eq:DeltaE}
\Delta E(x=i\lambda,t) = \langle \tilde{E}_i(t) - E^0_i(t) \rangle \ ,
\end{equ}
where $\tilde{E}_i(t)$ is  the energy contained in the $i$th  cell at time $t$
of   the   perturbed  trajectory   and   respectively   for  $E^0_i(t)$,   and
$\langle\cdot\rangle$  denotes  the  average  over an  ensemble  of  different
initial realizations.

\begin{figure}[!t]
\begin{center}
  \includegraphics[scale=0.47]{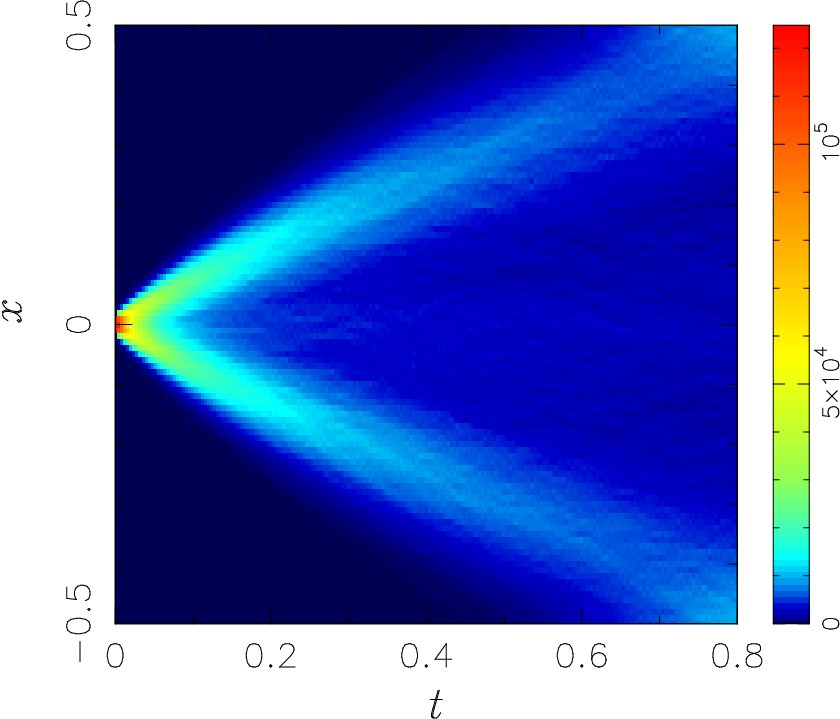}
  \includegraphics[scale=0.47]{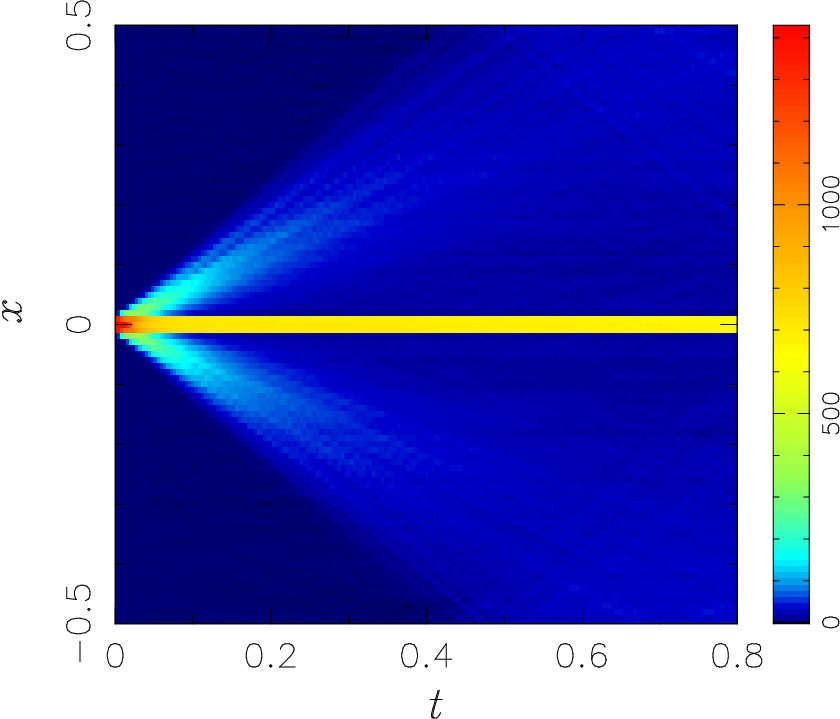}
\caption{Evolution of the energy difference $\Delta E(x,t)$  for a chain
  of $N=101$,  $n=50$ and  $\sigma=0.5$, and for  $\gamma=N$ (left  panel) and
  $\gamma=1$   (right   panel).    The   initial   energy   per
  degree of freedom   was
  $\varepsilon^0=5000$ and $\tilde{\varepsilon} = 50000$.
\label{fig:E-perturb}}
\end{center}
\end{figure}

When  dynamical  correlations  are  not   too  strong,  one  expects  that after
a sufficiently long time the perturbation $\Delta E(x,t)$ scales (with
$x$ measured from the initially perturbed cells)
as
\begin{equ} \label{eq:scaling}
\Delta E(x,t) = \frac{1}{t^\xi} \Delta E\left(\frac{x}{t^\xi},t\right) \ ,
\end{equ}
where the power $\xi$ is related to the scaling of the heat conductivity with
the size of the system $L$ as \cite{DKU2003}:
\begin{equ} \label{eq:kappa}
\kappa = N^{2 - 1/\xi} \ .
\end {equ}
In  particular, $\xi  = 1/2$  corresponds to  normal diffusion,  while $\xi=1$
corresponds to ballistic motion.

In  \fref{fig:E-perturb} we  show the  evolution of  the  energy difference $\Delta E(x,t)$ for the  deterministic chain $\gamma=N$ (left panel) and
compare  it with  the  stochastic  chain with  $\gamma=1$  (right panel).   We
observe that for $\gamma=N$, the initial  excess of energy at the central cell
decays very  rapidly.  Practically none of  the initial energy  remains in the
central cell.   The perturbation  moves symmetrically to the ends  of the
chain, carried by two seemingly  independent families of particles, those with
positive velocity  and those  with negative velocity.   Actually, it is  the fast
decay of the energy at the center that  marks the existence  of very
strong dynamical correlations. A similar observation has been made recently in
a random walk with memory in  the waiting times of successive steps \cite{Za06}.
For  $\gamma=1$,  we  also observe  an  initial  fast  decrease of  the  energy
perturbation. The evolution of the peaks with positive and negative velocity
seems to move ballistically. On the other hand, when $\gamma=N$, the
group velocity of the
outgoing peaks  depends weakly on  time, probably reaching a  final constant
velocity at much longer times.

\begin{figure}[!t]
\begin{center}
  \includegraphics[scale=1]{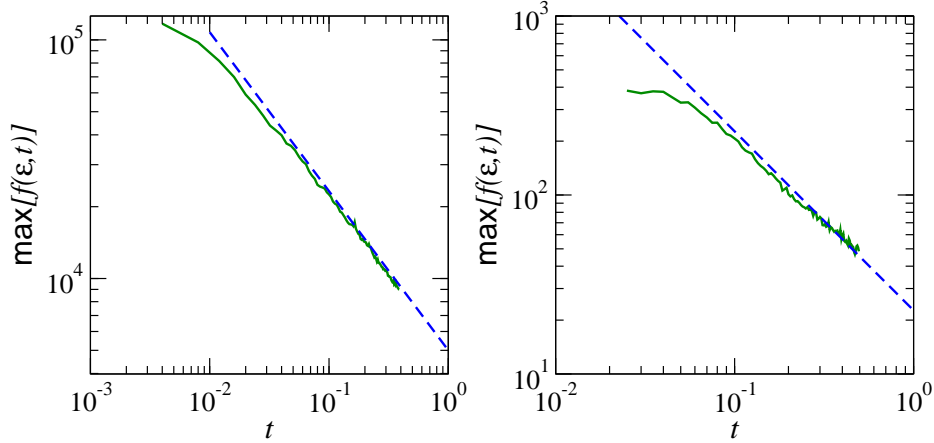}
\caption{Damping  of the maximum  value of  $\Delta E(x,t)$  as a  function of
  time, for the same simulation as in \fref{fig:E-perturb}, for $\gamma=N$
  (left panel)  and $\gamma=1$ (right panel).  The  dashed curves corresponds
  to the power law $~t^{-2/3}$ for $\gamma=N$ and to $~t^{-1}$ for $\gamma=1$.
\label{fig:damping-N}}
\end{center}
\end{figure}

The  scalings \eref{eq:scaling}  and  \eref{eq:kappa}  are  valid for  the
decaying of  the initial perturbation, namely  they are valid  if measured from
the decay of the central peak. Nevertheless, we find that in our case, similar
scalings  are possible for  the outgoing  peaks. Assuming  that the  excess of
energy is  transported across the  system as a  density packet, whose  area is
preserved on  average, we show  in \fref{fig:damping-N} the damping  of the
amplitude  of the  moving peak  as a  function of  time.
For $\gamma=N$  (left panel),  the amplitude  of the
peak decays as  $~t^{-2/3}$, corresponding to a heat  conductivity that scales
as $\kappa\approx N^{1/2}$. Note,  however, that within numerical accuracy,
the decay could also be consistent with $\kappa\approx N^{1/3}$ as it is found
at the beginning of this section.  In fact,  from  a  fit  to a power law  of  the
amplitude decay we have obtained scalings for the amplitude decay between
$0.62$ to  $0.66$. In any case,  the spreading of the  outgoing peak reflects,
locally,  the  anomalous  character  of  the  heat  transport.   On
the other hand,  the
amplitude decay for $\gamma=1$ (right panel)  is consistent with $~t^{-1}$, corresponding to
a heat conductivity  that grows linearly with $N$, in  full agreement with the
observations at the beginning of this section.

\begin{acknowledge}
  We thank E. Hairer for substantial help with the programming of
  Eq.~\eref{boltzmann}, and S. Lepri and F. Piazza for useful discussions. This work was partially supported by the Fonds
  National Suisse.
\end{acknowledge}

\bibliographystyle{unsrt}
\bibliography{balancier-9}

\end{document}